\documentstyle[aps,pre,tighten]{revtex}

\begin{document}

%%% \volnumpagesyear{0}{0}{000--000}{2005}
%%% \dates{received date}{revised date}{accepted date}

%\baselineskip=24pt
%** \baselineskip=16pt
%\baselineskip=12pt
%** \hoffset= -2.5mm

%\title{Dissipation, Transport, Noise:
%Mesoscopics beyond the Landauer-B\"uttiker Theory}
\title{Conservation, Dissipation, and Ballistics:
Mesoscopic Physics beyond\\
the Landauer-B\"uttiker Theory}

\author{Frederick Green and Mukunda P. Das}
\address{
Department of Theoretical Physics,
Institute of Advanced Studies,\\
Research School of Physical Sciences and Engineering,
The Australian National University,\\
Canberra ACT 0200, Australia.}
%\mailingone{fgreen@phys.unsw.edu.au}

%\author{Mukunda P. Das}
%\address{
%Department of Theoretical Physics,
%Institute of Advanced Studies,\\
%Research School of Physical Sciences and Engineering,
%The Australian National University,\\
%ACT 0200, Australia.}
%\mailingtwo{mukunda.das@anu.edu.au}

\maketitle

\begin{abstract}
The standard physical model of contemporary mesoscopic noise and
transport consists in a phenomenologically based approach, proposed
originally by Landauer and since continued and amplified by B\"uttiker
and others. Throughout all the years of its gestation and growth,
it is surprising that the Landauer-B\"uttiker approach to mesoscopics
has matured with scant attention to the conservation properties lying at
its roots: that is, at the level of actual microscopic principles.
We systematically apply the conserving sum rules for the electron gas
to clarify this fundamental issue within the standard phenomenology
of mesoscopic conduction. Noise, as observed in quantum point contacts,
provides the vital clue.

%\smallskip
%\noindent
%{Keywords: {\em ballistic transport; nonequilibrium noise; quantum kinetics;
%conservation laws; sum rules; quantum point contacts}}
\end{abstract}

\section{Introduction}

\subsection{Background}

If someone were to declare that a successful kinetic theory is one that
explains experiment without any need to obey Newton's laws, such an
assertion would not be likely to be received unreservedly -- even
by most philosophers of science. In this paper we ask what it would
mean if, in some sense, an analogous proposition were to have been made
to a segment of the mesoscopics community, and implicitly accepted.
It would be a curious situation indeed; one with intriguing implications
for the pursuit of mesoscopic physics and, in particular, for noise theory.

Here, the conservation laws of charge and particle number stand
at stage center. The principles of conservation are in every sense
as cardinal as Newton's laws. In transport physics, their validity
is neither compliant to any quantitative revision nor contingent
upon any set of intuitive preconceptions. Conservation is the
inalienable heart of every rational model of conduction.
There is no exception.

%To set the background for our investigation, let us revisit
Let us revisit
the essentials of transport in mesoscopic conductors. Typically,
structures have lengths comparable to, or smaller than,
the mean free paths for electron scattering in the bulk. Charge
transport can then be regarded as ballistic (collision-free)
within the confines of such a metallic channel.
\footnote{The term ``metallic channel'' has the specific
sense of a device with a band of extended conducting states
that can be freely populated and depopulated by controlling the carrier
density. Carriers have free access to and from a set of (large)
bounding leads which are themselves strongly metallic.}

\vskip 0.5truecm
%%%-- 
\input psfig.sty
%%%-- 
\centerline{\hskip10mm\psfig{figure=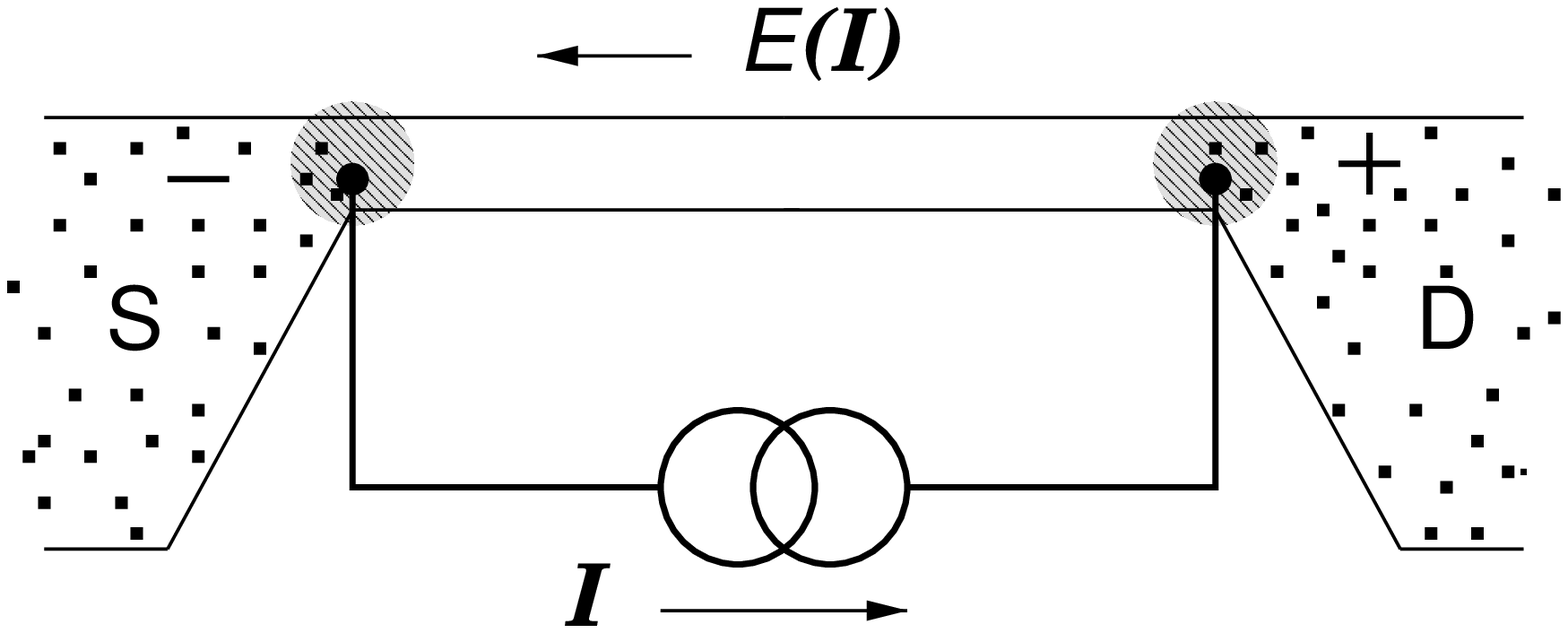,height=3.5truecm}}
\vskip 0.5 truecm
%%%-- 
\noindent
%%%-- 
{\bf Figure 1.}
%%\caption
{\small
An ideal, uniform ballistic wire, or quantum point contact.
Its diffusive and dissipative leads (S and D, stippled) are permanently
neutral and permanently in equilibrium. A paired source and sink
of current $I$ at the boundaries drive the transport.
Local charge clouds (shaded),
induced by the imposed influx and efflux of $I$, set up the dipole
voltage $V = E(I)L$ between D and S. The ballistic wire manifests
its optimal (Landauer) conductance $I/V$ if, and only if, the
mean free paths for elastic and inelastic scattering are equally
matched and straddle the whole wire, from its strongly dissipative
source terminal to the corresponding drain.
See Section 3 for a detailed proof.
} 
%\end{figure}
\vskip 0.5 truecm

Figure 1 illustrates this
for a so-called ``quantum point contact'' (QPC).
We are dealing with small -- potentially quasi-molecular -- structures.
Consequently they experience a high degree of {\it openness} to their
macroscopic environment. The restricted dimensionality also
accentuates the inter-particle {\it correlations}. It is no secret
\cite{[9],[0]}
that these are the very attributes that hold the hardest set
of challenges to be met in developing mesoscopic electronics.

Among the challenges, the foremost is how to reconcile
{\it microscopic conservation} with both openness at the device
interfaces and strong quantum correlations just about everywhere else.
Without a firm anchoring in the conservation laws, nothing
prevents an unconstrained model from ``creating'' or ``destroying''
charges and currents willy-nilly. Do we need such a model?

Even if seemingly able to explain an experiment, a theory that is
nonconserving must be regarded as fundamentally incoherent, for
we have learned that Nature does not work in this way.
That is the case, anyway, in the realm of condensed matter.
The absolute centrality of conservation in mesoscopic transport
needs no further comment.

\subsection{Plan of the Paper}

The paper is set up as follows. In the next Section we look at the
prime relation of mesoscopic transport; that is, the immediate link
between finite conductance and finite energy dissipation.
This leads into the third Section and our first cardinal result:
the strictly kinetic derivation of Landauer's quantized conductance
formula for an ideal metallic channel
\cite{[0],[1],[1a]},
using only the axioms of microscopic theory and free of any recourse
to the standard model's very special assumptions. The message is
that a mesoscopic model (one which is faithful to orthodox
quantum kinetics) does not need phenomenology for a backbone.

Our result shows that the Landauer-B\"uttiker (LB) project
is already subsumed in a more mature, fundamentally established,
and truly first-principles framework: quantum kinetics. This means that
the phenomenological assumptions, on which the unique mesoscopic
character of LB has been predicated
\cite{[0]},
are not foundational.

Later in Sec. 3, then in 4, we go on to demonstrate how the logic
of the conservation laws generates a completely natural
microscopic description of a major exemplar of
mesoscopic fluctuations: QPC noise.
We account for data that went unexplained for nearly a decade. In the
process it becomes clearer that the standard
Landauer-B\"uttiker approach
to fluctuations does not respect conservation. Because of that, it
fails to predict the directly observed consequences of conservation.

Finally we sum up the origin, motivation, and cogency of our
microscopically conserving analysis. We also retrace the implications
for setting up a novel perspective on mesoscopic processes.
Interested readers will find an Appendix containing the formal
proof that the Landauer-B\"uttiker noise theory
does not satisfy the {\it perfect-screening} and
{\it compressibility} sum rules. These basic rules govern the
dynamics of mesoscopic noise, as experimentally demonstrated for QPCs.

\section{The Physical Issue}
%\vskip 12 truept

At the outset, we remark that our analysis applies specifically to
{\em metallic} conduction and fluctuations. We examine open mesoscopic
structures where the carrier states form a quasi-continuum, are
spatially extended, and have considerable overlap. This is territory
that the LB approach is supposed to cover extremely well
\cite{[0]}.
We are not considering -- here -- systems in which quite different,
nonmetallic mechanisms of transport (e.g. hopping and tunneling)
involve electronic orbitals that are discrete, highly localized,
and with little overlap.

A simple intuitive understanding of mesoscopic transport and noise
was established, in the last fifteen years or so, through the
insights of Landauer, B\"uttiker, Imry, and many similarly inclined
contributors
\cite{[0],[1],[1a],[2]}.
(The sheer proliferation of such works precludes citing
all but a few outstanding examples.)
In that succinct perspective the physical basis of current
flow, mesoscopic and otherwise, is identified as a mismatch of
carrier density between two or more metallic reservoirs -- terminals --
across which the QPC or other low-dimensional conductor
%of interest
is connected.

The hallmark of metallic transport in quantum point contacts,
%namely
the quantization of conductance as ``Landauer steps'' in units of
$2e^2/h$ ($\approx$ 0.078 mS), appears to be adequately
explained in terms of coherent transmission of electron waves
through a perfectly loss-free barrier. However,
%the picture of quantum-coherent scattering at this simple level
a picture of quantum-coherent scattering as simple as this
cannot address the central theoretical issue of metallic conduction:

\vskip 8 truept
\centerline{
{\it What causes dissipation in a ballistic quantum point contact}?}
\vskip 8 truept

\noindent
Such a question is more than academic. In the near future,
reliable and effective nano-electronic design will demand models
that are credible; credible not just as physical theory but
as engineering practice. Therefore a comprehensive theory has to
confront the unavoidable presence, and action, of power dissipation.
%\eject

For conduction, the central issue is plain. Any
finite conductance $G$ must {\it dissipate} electrical energy
(inelastically) at the rate
$P = IV = GV^2$, where $I = GV$ is the current and $V$ the potential
difference across the terminals of the driven channel.
This is mandated by the underlying physics of charge conservation
in externally driven (open) conduction
\cite{[7],[8]};
as such it cannot be hostage to any transport
philosophy insisting on coherent propagation
as the exclusive origin of conductance.

Our question has its definitive answer in many-body quantum kinetics
\cite{[3]},
free of all ``supplementary hand-waving''
\cite{[0]}.
The application of well tested microscopic methods leads not only
to conductance quantization precisely by the direct inclusion
of inelastic energy loss
\cite{[8a],[3],[4],[4a]},
but, as we show in Sec. 3 to follow, it also resolves a
long-standing experimental enigma
\cite{[5]}
in the noise spectrum of a quantum point contact (QPC)
\cite{[6]}.
The same developments foreshadow a systematic pathway to
truly predictive design of novel structures.

In their dissipative action, inelastic collisions are beyond the reach 
of transport models that rely on coherent quantum scattering alone
to explain the origin of $G$. Coherence
implies elasticity, and elastic scattering is always loss-free:
it conserves the energy of the scattered particle. The recent review
of mesoscopic phenomenology by Agra\"\i t and others
\cite{[9]}
expresses a similar conclusion on the insufficiency of
coherence-based arguments for actual experimental systems.

There is no choice but to explicitly admit the
dissipative mechanisms (e.g. phonon emission) by which the
energy gained by carriers, when transported from source to drain,
is returned incoherently to the surroundings. A very detailed
microscopic demonstration of this has been given by Sor\'ee {\em et al.}
\cite{[8a]}.
Thus, alongside elastic and coherent scattering,
inelastic processes are always in place.
Together, they fix $G$; yet it is only the energy-dissipating
mechanisms that secure {\it thermodynamic stability} in
steady-state conduction. Coherence on its own cannot bring this about.

A serious deficiency underlies purely elastic approaches to transmission.
They cannot handle dissipation.
We now review the well-defined microscopic resolution of that problem.

\section{The Physical Solution}

The complete microscopic understanding
of the ubiquitous power-loss formula $P = GV^2$
\cite{[7],[8],[10]}
is couched in terms of the fluctuation-dissipation theorem
(details are standard in the kinetic literature
\cite{[10],[10a]}).
It holds for {\it all} resistive devices at all scales
without exception.
The theorem expresses
the condition of thermodynamic stability.
\footnote{There is an intimate structural correspondence
between the dissipation formula $P = IV$, originating from
the canonical, many-body Hamiltonian description of a driven conductor 
\cite{[8]},
and the form of the steady-state solution to the
quantum kinetic equation for its carrier distribution.
In the latter case, the fluctuation-dissipation theorem
follows directly when the limit of a weak driving field is taken.
The formalities are in Ref.
\cite{[4a]},
which also shows the fundamental role played by
{\em electron-hole polarization} in determining the
current-current correlator and the conductance.}
With it comes the conclusion that
\cite{[3],[4],[4a]}
\vskip 8 truept

{$\bullet$} 
inelasticity is necessary and sufficient to stabilize current
flow at finite conductance;

{$\bullet$} ballistic quantum point contacts have finite
$G \propto 2e^2/h$; therefore

{$\bullet$} processes of energy loss are indispensable
to any rational theory of ballistic transport.
\vskip 8 truept

To allow for the energy dissipation vital to any microscopic
description of ballistic transport, we recall that open-boundary
conditions imply the intimate coupling of the QPC channel to its interfaces
with the reservoirs. The interface regions must be treated
as an integral part of the device model. They are the actual sites
for strong scattering effects: {\it dissipative} many-body events
as the current  enters and leaves the ballistic channel,
and {\it elastic} one-body events as the carriers interact
with background impurities, the potential barriers that confine and funnel
the current, and so on. See Fig. 1 for an illustration.

A key idea in our treatment is to subsume the interfaces within
the total kinetic description of the ballistic channel.
At the same time, strict charge conservation in an {\it open}
device requires the direct supply and removal of charge flux
by a generator outside the system of interest
\cite{[7]}.
With this canonical precondition -- that an open channel
must be energized by a strictly {\it external} driving
agent -- the current is completely independent of the physics
locally governed by the reservoirs.

That fundamental condition sets the
quantum kinetic approach totally apart from Landauer-like treatments
\cite{[0],[1],[1a],[2]}.
Instead of the gauge-invariant open-system requirement
\cite{[7],[8]}
that any current flow is perforce supplied independently from outside,
\footnote{Transport at fixed current and at fixed voltage
are completely equivalent. The electrodynamics of conduction
in an open-boundary device driven by an external current generator
\cite{[7]}
is isomorphic to that of a closed conducting loop, comprising
the ``open'' device and a voltage source (a battery)
\cite{[8]}.
Discussion of the underlying gauge invariance is beyond
our scope here. We urge the interested reader to consult,
in parallel, both Sols 
\cite{[7]}
and Magnus \& Schoenmaker
\cite{[8]}.}
they rely on the intuitive assumption that the current depends,
passively, on a notional density ``mismatch'' between reservoirs.
It is nothing other than an {\it ad hoc} extrapolation,
to strongly quantum systems, of purely classical diffusion
\cite{[1a]}.

%\vskip 8 truept
%\centerline{{\it a. Ballistic Conductance}}
%\vskip 8 truept
\subsection{Ballistic Conductance}

One can straightforwardly write the algebra for the conductance
in our model system. A uniform, one-dimensional ballistic QPC, of
operational length $L$, will be associated with two mean free paths
determined by $v_{\rm F}$, the Fermi velocity of the electrons,
and a pair of characteristic scattering times
$\tau_{\rm el},\tau_{\rm in}$. Thus

\begin{equation}
\lambda_{\rm el} = v_{\rm F}\tau_{\rm el};{~~}
\lambda_{\rm in} = v_{\rm F}\tau_{\rm in}.
\label{e1}
%\eqno(1)
\end{equation}

\noindent
Respectively, these are the scattering lengths set by the
elastic and inelastic processes active at both interfaces. The
device (i.e. the QPC {\it with} its interfaces) has a conductive
core that is collisionless. It follows that

\begin{equation}
\lambda_{\rm el} = L = \lambda_{\rm in}.
\label{e2}
%\eqno(2)
\end{equation}

Finally, the channel's conductance is given by the familiar formula
(deducible straight from Kubo's microscopic prescription
\cite{[3],[10]})

\begin{equation}
G = {ne^2\tau_{\rm tot}\over m^*L}
= {2k_{\rm F}\over \pi}{e^2\over m^*L}
{\left( {\tau_{\rm in}\tau_{\rm el} \over
         {\tau_{\rm el} + \tau_{\rm in}}} \right)};
%\eqno(3)
\label{e3}
\end{equation}

%\eject
\noindent
the effective mass of the carriers is $m^*$.
The first factor of the rightmost expression for $G$
has the density $n$ recast in terms of the Fermi momentum $k_{\rm F}$;
in the final factor, Matthiessen's rule
$\tau_{\rm tot}^{-1} = \tau_{\rm el}^{-1} + \tau_{\rm in}^{-1}$
fixes the total scattering rate.

On applying Equations (1)--(3), the conductance reduces to

%\eject

\begin{equation}
G = 2{e^2\over \pi\hbar}{\hbar k_{\rm F}\over m^* L}
{\left( {(L/v_{\rm F})^2 \over 2L/v_{\rm F}} \right)}
= {2e^2\over h} \equiv G_0.
%\eqno(4)
\label{e4}
\end{equation}

\noindent
This is no more -- and no less -- than the Landauer conductance of a
single, one-dimensional, ideal metallic channel.

What is the most important lesson that one can draw
from this microscopic result?
It is that Occam's Razor can always be used to advantage.
Of two contending explanations of a phenomenon, choose
the one with fewer hypotheses.

The microscopic interpretation of the Landauer formula
{\it does not need any} of the adventitious assumptions
otherwise invoked to explain conductance quantization
\cite{[0],[1],[1a],[2]}.
The tight derivation of Eq. (4) follows from completely standard
quantum kinetics for open systems, within which total phase
coherence is not even a physical possibility
let alone a theoretical necessity. Nor are the remaining,
singular assumptions of the Landauer-B\"uttiker
approach relevant to actual ballistic transport,
any more than coherence is. See Ref.
%\online
\cite{[3]}
for further discussion.

In Figure 2 we plot the results of our model for a QPC
made up of two one-dimensional conduction bands at energies
5$k_{\rm B}T$ and 17$k_{\rm B}T$, in thermal units at temperature $T$.
We use the natural extension of Eq. (4) to cases where
one or more channels may be open to conduction
\cite{[3]},
depending on $T$
and the size of the chemical potential $\mu$. As the role of
inelastic scattering is enhanced ($\tau_{\rm in} < \tau_{\rm el}$)
the conductance deviates from the ideally ballistic Landauer limit.

At the core of the quantum-kinetic Landauer formula
is the clear and pivotal influence of inelastic energy loss.
It is one of the underpinnings of quantum transport.
Charge conservation, the complementary underpinning, is guaranteed
by the use of microscopically consistent open-boundary conditions
at the interfaces
\cite{[7],[8]}.

%%\begin{figure}[htbp] 
%\vskip -12 truept
%\vskip 0.5 truecm
%%\centering{\resizebox{7.5cm}{!}{\includegraphics{appc9_dg_fig1.ps}}}
\input psfig.sty
\centerline{\hskip10mm\psfig{figure=appc9_dg_fig1.ps,height=7.0truecm}}
%\centering{\frame{\rule{3cm}{3cm}}}
%\centering{\frame{\rule{5cm}{5cm}}}
%\vskip 0.5 truecm
\noindent
{\bf Figure 2.}
%%\caption
{\small
Conductance quantization in a two-band
ballistic point contact, as a function of chemical
potential $\mu$, calculated from Eq. (\ref{e3})
within our conserving kinetic theory; see Ref.
%\online
\cite{[3]}.
Full curve: ideal ballistic
channels. Broken curves: non-ideal behavior increases
with the onset of inelastic phonon emission inside the contact,
which preserves the Landauer steps while suppressing their height.
}
%\end{figure}
%
\vskip 0.5 truecm

The twin requirements of open-system dissipation and
open-system gauge invariance are
neither adequately considered nor, as it turns out, respected by
any of the phenomenological derivations of Eq. (4)
\cite{[0],[1],[1a],[2]}.
If this is so at the plain level of
conduction, such approaches will be far more
problematic for current fluctuations,
a decisive test of microscopic transport theories.
We discuss that very point now.

%\vskip 8 truept
%\centerline{{\it b. Non-equilibrium Noise}}
%\vskip 8 truept
\subsection{Nonequilibrium Noise}

The noise response of a quantum point contact is a fascinating
aspect of mesoscopic transport. It is a much more demanding one both
experimentally and theoretically.
%\noindent
In 1995, a landmark measurement
of nonequilibrium noise was performed by the Weizmann group
\cite{[5]},
which yielded a tantalizing result. While conventional models
\cite{[2]}
predict no fine structure {\it whatsoever} in the noise of
a QPC driven at constant current levels, the measured data
(Fig. 3, left-hand panel) show a prominent and robust set of
peaks at threshold, just where the carrier density in the QPC
starts to enter the degenerate metallic regime.

%\vskip 0.5 truecm
%%\begin{figure}[htbp] 
%
%%\centering{\resizebox{13.5cm}{!}{\includegraphics{appc9_dg_fig2.ps}}}
\input psfig.sty
\centerline{\hskip10mm\psfig{figure=appc9_dg_fig2.ps,height=6.0truecm}}
%\centering{\frame{\rule{3cm}{3cm}}}
%\centering{\frame{\rule{5cm}{5cm}}}
%\vskip -3truemm
\noindent
{\bf Figure 3.}
%\caption
{\small
Nonequilibrium low-frequency current noise of a QPC at constant
source-drain current, as a function of gate bias. Left: the data
from Reznikov {\it et al.}, Ref.
%\online
\cite{[5]},
display a dramatic series of peaks coincident with the energy
threshold of the lowest subband, where carriers first populate
the states for conduction.
Right: calculation
from Green {\it et al.}, Ref.
%\online
\cite{[6]}.
In each case the dotted line at 100nA shows
the shot-noise prediction of the Landauer-B\"uttiker
(LB) formula
\cite{[2]},
using as its required phenomenological inputs
the measured and calculated data for $G$.
The LB formula falls well short,
qualitatively and quantitatively;
compare the dash-dotted lines also at $I = 100$nA.
}
%\end{figure}
\vskip 0.5 truecm

We have accounted for the Reznikov
{\it et al.} measurements within a strictly conserving, kinetic
formulation of nonequilibrium noise in a quantum point contact
\cite{[6]}.
Figure 3 displays, side by side with the experimental data,
our computation of excess QPC noise under the same operating
conditions.

In contrast to the outcome of popular mesoscopic phenomenology
\cite{[2]}
one notes the close affinity between the measurements and the quantum
kinetic calculation,
as the carrier density is swept across the first conduction-band
threshold, where the conductance exhibits its lowest step.
At fixed values of source-drain current, the accepted noise account
\cite{[2]}
predicts no peaks at all, but rather a featureless monotonic drop
in the noise strength as the carrier density passes through threshold.

Remarkable as they are to this day, the Weizmann measurements have
remained absolutely unexplained for a decade. Moreover their obvious
message, namely that established theories are inadequate to the
experiment, was simply ignored by the mesoscopic community.
To the contrary, the folklore seemed to spread that Ref.
%\online
\cite{[5]}
amply confirmed accepted understandings
\cite{[2]}.

\section{The Power of the Conservation Laws}

The key to all quantum kinetic descriptions of conductance is the
fluctuation-dissipation theorem
\cite{[10a]},
whose practical implementation is
Eq. (3) (where the overall relaxation time $\tau_{\rm tot}$
encodes all the electron-fluctuation dynamics via the Kubo formula
\cite{[10]}).
This universal relation is one of the electron-gas
{\it sum rules}
\cite{[11]}.
In this instance, it expresses the conservation of energy, dissipatively
transferred from an external source to the thermal surroundings, for
any process that involves resistive transport -- including
that in a ballistic quantum point contact.

A second, and equally fundamental, sum rule concerns the compressibility
of an electron fluid in a conductive channel. This sum rule turns out
to have an immediate link with the nonequilibrium noise behavior
reviewed above. Here we give a brief explanation of that crucial link.
For all of the formal details, see Refs.
%\online
\cite{[6],[6a],[12],[14]}.

Recall that the carriers in a metallic quantum point contact are stabilized
by the presence of the large leads, which pin the electron density
to fixed values on the outer boundaries of the interfaces (recall too
that the interfaces and the channel together define the open system).
No matter what the transport processes within the QPC may be,
or how extreme, the system's {\it global neutrality} is guaranteed
by the stability of the large and charge-neutral reservoirs.
\footnote{
Independently of Sols' theorem
\cite{[7]}
and its consequences, the importance of global neutrality
and metallic screening by the reservoirs was driven home long ago,
in great microscopic detail, by Fenton
\cite{[14a]}.
His analysis shows that it is confinement of the electric field within
the QPC (via perfect screening), and {\it not} perfect elastic
transmission, that is a crucial prerequisite for conductance quantization.} 

It follows that the mean total number $N$ of active electrons in the
device remains independent of any current that is forced
through the channel, for $N$ is always neutralized by the rigid ionic
background in its neighborhood as well as the stabilizing leads.
The presence of the latter means that any and all fringing fields
are screened out beyond the device boundaries;
hence the global neutrality.

%\eject
It is readily seen that, if the mean occupation number $N(\mu)$ is
independent of any external current as a result of global neutrality
(i.e. the perfect-screening sum rule
\cite{[11]}),
so is the total mean-square number fluctuation
$\Delta N = k_{\rm B}T{\partial N/\partial \mu}$
\cite{[12]}.
The carriers' compressibility in the QPC is specified
in terms of $N$ and $\Delta N$ by
\cite{[11]}

\begin{equation}
\kappa = \kappa_{\rm cl} {\Delta N\over N}
{~~}{\rm where~~}\kappa_{\rm cl} \equiv {L\over Nk_{\rm B}T},
%\eqno(5)
\label{e5}
\end{equation}

\noindent
which, in consequence, remains strictly unaffected by any transport process.

This a surprising corollary of global neutrality. It asserts that, in
an open conductor, the system's equilibrium compressibility completely
determines the compressibility of the electrons even away from
equilibrium, {\it regardless} of the strength of the driving field.

The compressibility sum rule expresses the unconditional
conservation of carriers in a nonequilibrium conductor.
Previously unexamined in mesoscopics, this principle has
an immediate importance and applicability. 
It means that a new, independent, and strong criterion has become
available to test the consistency of any model for ballistic transport. 

How does $\kappa$ determine the noise in a QPC? The strength
of the current fluctuations is, in broad physical terms, a convolution
of two competing phenomena:

\begin{equation}
S(I,t) \sim {\langle I(t)I(0) \rangle}
{\Delta N\over N}
= {\langle I(t)I(0) \rangle} {\kappa\over \kappa_{\rm cl}}.
%\eqno(6)
\label{e6}
\end{equation}

\noindent
The leading factor represents the self-correlation of the
dynamical electron current $I(t)$ evaluated as a trace
over the nonequilibrium distribution of excited
electrons throughout the device region. The second factor -- evidently
a basic characteristic of the electron gas in the channel -- is
independent of $I$, meaning that the invariant compressibility,
Eq. (5), dictates the overall scale of the noise spectrum.

Let us analyze the noise curves of Fig. 3 in light of this
microscopic result
\cite{[6],[6a]}.

\begin{itemize}
\item
% {$\bullet$} 
At large negative bias $V_g$, the channel is depleted.
The remnant
carriers are classical, so $\Delta N/N \to 1$. The noise is then
dominated by strong inelastic processes at high driving fields,
represented within the structure ${\langle I(t)I(0) \rangle}$.
\item
% {$\bullet$} 
In the opposite bias limit (right-hand sector of each panel
in Fig. 3), the channel is richly populated and thus
highly degenerate, with a large Fermi energy $E_{\rm F}$.
Then $\Delta N/N \to k_{\rm B}T/2E_{\rm F} \ll 1$. The noise
spectrum falls off according to Eq. (6), since the
current-correlation factor -- now comfortably within the regime of
ballistic operation -- reaches a fixed ideal value.
\item
% {$\bullet$} 
In the mid-range of $V_g$ there is a point where the
carriers' chemical potential matches the energy threshold for
populating the first conduction sub-band. Here there is robust
competition. As inelastic scattering becomes less effective,
the correlation ${\langle I(t)I(0) \rangle}$ grows
while the onset of degeneracy starts to cut down the
other factor, the compressibility.
Where this interplay is strongest, peaks appear.
\end{itemize}

\noindent
The outworking of the compressibility rule is clear: it is,
quite directly, the ``inexplicable'' emergence of the noise peak structures.
The striking case of QPC noise gives an insight into the
central importance of the conserving sum rules in the physics
of metallic transport at meso- and nanoscopic dimensions.

As detailed in the Appendix, the more phenomenological treatments
of noise fail to address the explicit action of microscopic conservation
in ballistic phenomena, to the point that control over the sum-rule
violations does not exist for them.
Therefore they cannot offer a rational understanding
of the real nature of ballistic conduction -- much less its noise. 

\section{Summary}

The kinetic analysis of transport provides a detailed,
fully microscopic account of conductance and noise. This applies to
the specific case of quantum point contacts. We accurately reproduce
the entire current response of mesoscopic conductors, in particular
conductance quantization. The keys to this newly fruitful picture
are {\it open-system charge conservation} and
the physical reality of {\it dissipative scattering}.

Our unified theory yields a comprehensive understanding of the
nonequilibrium fluctuations fundamental to a QPC. We have successfully
tested this comprehension by fully explaining the long-standing puzzle
posed by the noise measurements of Reznikov {\it et al.}
\cite{[5]}.
The theoretical impact of noise and fluctuation physics is that it
carries much more information on the internal dynamics of mesoscopic
systems -- knowledge that is not accessible through the $I$-$V$
characteristics alone.

The capacity to advance a self-contained microscopic explanation
of processes in mesoscopic transport underwrites a matching
ability to build up a program for device design that is 
inherently rational. This goes deeper than assembling,
{\it ad hoc}, any clutch
of ideas that happen to fall easily to hand. The reward for more
workmanlike efforts is a huge potential to improve the
practical engineering of novel generations of devices.
These can -- and evidently should -- be built
on a sound and non-speculative knowledge base.
Many-body physics is at the heart of the program.

A well grounded mesoscopic theory, at its simplest,
will not be simplistic.
To go so far as to forget or even ignore the conservation laws,
merely to fit subjective notions of ``simplicity'',
would scarcely be science in the tradition
of Boltzmann, Sommerfeld, and Landau.
At any rate, it would not do for serious investigations
of co-ordinated behavior in matter at small scales.
For, a nonconserving model is incoherent;
and an incoherent model is unusable.
Nor does the continued shirking of honest and {\em public} debate
on the issue serve the credibility of mesoscopic theory.

It is the universal principles of conservation
that are the gatekeepers to a real mesoscopic understanding.
We have revisited the conserving sum rules in their specific
application to small-scale metallic transport. Given its prime
role in carrier dynamics, a truly conservative quantum kinetic
approach holds opportunities for both traditional many-particle
theory applied to mesoscopics, and for expanding noise research.
More than ever, the spotlight falls on the {\it fluctuation}
properties of active mesoscopic devices. Noise physics is
vital not only for its own sake, but as a robust and
trustworthy diagnostic means for electronic concepts to come.

Now is a good time indeed to introduce genuine many-body
methods into mesoscopics: a field in no small need of a
microscopically cohesive, and more far-reaching, vision.
Some of the tools to make a start on this are freely available
\cite{[7],[8],[8a],[3],[4],[4a],[6],[6a],[10],[10a],[11],[12],[14],[14a]}.

\section*{Acknowledgments}

Many friends have encouraged us consistently throughout our
fluctuation work. First we thank Jagdish Thakur, our
collaborator on the QPC noise and compressibility problems.
F.G. also recalls the strong and unstinting support of
Neil Barrett and Grant Griffiths, whose foresight first
empowered him to pursue this line of research.

Our review would be incomplete without recognizing the pioneering
contributions of Ed Fenton. Finally we acknowledge the
fruitful exchanges we have had with Fernando Sols, Wim Magnus,
Wim Schoenmaker, and Bart Sor\'ee.

\appendix
\section{Sum Rules in the Landauer-B\"uttiker Theory}

\vskip -8 truept
\noindent
{\it ``To lose one parent, Mr Worthing, may be regarded as a misfortune;
to lose both looks like carelessness.''} Oscar Wilde
\vskip 6 truept

In this Appendix we provide, for the record, the formal demonstration
that the Landauer-B\"uttiker model
violates both crucial sum rules for perfect screening (charge conservation)
and compressibility (particle conservation)
\cite{[11],[12],[14]}.
We are not by any means the first to examine the defective
gauge invariance of this phenomenology. The problem is already
well understood and discussed, for example,
by Blanter and B\"uttiker, Ref.
\onlinecite{[2]}.
See their Eq. (51) {\em et seq.}.

\subsection{Perfect Screening Sum Rule}

As we saw in Sec. 4, the overall neutrality of a quantum point
contact (indeed, of any bounded conductor) is maintained by the
strong screening property of its large conductive leads.
Suppose the QPC sustains a carrier density profile $n(x)$.
When $n$ is integrated for the total carrier number,
the contact's global neutrality implies

\vskip -8 truept
\begin{equation}
\int^L_0 n(x)dx~ \equiv N = {\rm constant}; ~{\rm that~ is,}~
{\partial N\over \partial I} = 0 {~~}{\rm for~ all~} I.
\label{A1}
\end{equation}
\vskip -1 truemm

\noindent
Regardless of external current $I$ forced through the structure,
and the voltage $V$ measured across it in response, the total of
mobile charges dwelling in the QPC (including the active interface
regions themselves) cannot depend on $I$ or $V$.
The entire structure, though open, stays unconditionally neutral.
This is the practical meaning of perfect screening.

Now consider the same situation from the perspective of
Landauer and B\"uttiker
\cite{[0],[1],[1a],[2]}.
In place of the gauge-invariant (and thus microscopically
correct) prescription of open, externally driven transport
that leads to perfect screening
\cite{[7],[8],[14]},
we posit that the current in a narrow conducting wire
is sustained purely by a difference of chemical potentials.
\footnote{One of the central phenomenological assumptions of LB
is that there must be a difference of chemical potentials between
the leads of a driven device. Otherwise, in LB, the current cannot
flow ``diffusively''
\cite{[1a]}
This phenomenology has absolutely no basis
in the quantum kinetic description of mesoscopic transport.
For, it would mean that $\mu$ -- a thermodynamic property which is
{\it always locally invariant} and guarantees the individual stability of
each lead's neutral state -- is subject to arbitrary change.
In reality it is the external driving source,
and not any ``difference'' in $\mu$, that directly sustains the current.
Thus in kinetic theory the leads'
chemical potentials are left alone to do their proper job:
to stabilize the system and maintain its overall neutrality.}
between an upstream electron reservoir
(chemical potential $\mu = \mu_{\rm up}$)
and a different downstream one
($\mu = \mu_{\rm dn} = \mu_{\rm up} - eV$).

The density along the conductor would then be some
function $n(\mu_{\rm up} - eV(x))$ that took the
boundary values $n(\mu_{\rm up})$ and $n(\mu_{\rm dn})$
at the ends of the sample.
According to such an account, in the linear-response
limit the total number of carriers in the active structure
of length $L$ would {\it change}, with the field, by
\vskip -1 truemm

\begin{eqnarray}
N(V) - N(0)
&=& \int^L_0 [n(\mu_{\rm up} - eV(x)) - n(\mu_{\rm up})] dx
\cr
&\to& \int^L_0 {dn\over d\mu_{\rm up}} (-eV(x)) dx
\cr
&\approx& - \Delta N(0) {eV\over 2k_{\rm B}T} ~\ne~ 0,
\label{A2}
\end{eqnarray}
\vskip -1 truemm

\noindent
a result manifestly counter to
the perfect-screening sum rule, Eq. (\ref{A1}), and
thus to gauge invariance.

\subsection{Compressibility Sum Rule}

We remind ourselves that the corollary to perfect screening
is the nonequilibrium compressibility sum rule
(recall Eq. (\ref{e5})):

\begin{equation}
{\kappa\over \kappa_{\rm cl}} \equiv {k_{\rm B}T\over N}
{\partial N\over \partial \mu} = {\rm const.};~{\rm that~ is,}~~
{\partial \kappa\over \partial I} = 0 {~~}{\rm for~ all~} I.
\label{A3}
\end{equation}
%\vskip -1 truemm

\noindent
Note in particular that $\kappa$ is calculable
as a strictly {\it equilibrium}, and not a linear-response,
quantity
\cite{[11]}.
It cannot depend on any transport parameter.

How does compressibility relate to the so-called Landauer-B\"uttiker
noise formula? The formula emerges from quite a variety of different
arguments, which nevertheless all converge to the same final expression
(specializing to one subband will not alter our own argument):

\begin{equation}
{\cal S}(V)
= 4G_0k_{\rm B}T
{\left[ {\cal T}^2(v_{\rm F}) +
{\cal T}(v_{\rm F}){\Bigl( 1\!-\!{\cal T}(v_{\rm F}) \Bigr)}
{{\mu_{\rm up}\!-\!{\mu}_{\rm dn}}\over 2k_{\rm B}T}
{\rm coth}{\left(
{{\mu_{\rm up}\!-\!{\mu}_{\rm dn}}\over 2k_{\rm B}T}
\right)}
\right]}.
\label{A4.0}
\end{equation}
\vskip -1 truemm

\noindent
The coefficient ${\cal T}(v_{\rm F})$ for coherent transmission
is the probability that an incoming carrier at Fermi velocity $v_{\rm F}$
will propagate coherently from source to drain,
right through the subband in the QPC.

The noise formula is the sum of separate contributions
\cite{[2]}.
With these same components, one may also reconstruct
the LB version of the mean-square number fluctuation.
After constructing the would-be ``single-carrier'' fluctuation,
\footnote{In a charged Fermi gas the fluctuation that should be
averaged for $\Delta N$ is an {\it electron-hole pair} which is
correlated, internally and irreducibly, via conservation of
energy, momentum -- and charge
\cite{[11]}.
It {\it cannot} be expressed as a product of two stochastically
independent single-particle terms.}
say ${\rm d} N$ (refer to Martin and Landauer
\cite{[15]}
for a clear account of this methodology), 
we obtain the LB mean-square expectation $\Delta N$
over the quantum point contact:
\footnote
{Since in LB the active states are limited to lie within a (thin)
shell $k_{\rm B}T$ centered on the Fermi surface,
the velocity of all carriers contributing to $S(V)$ always has
magnitude ${\langle |v| \rangle} = v_{\rm F} \gg {\langle v \rangle}$.
Therefore the mean-square LB fluctuation in one dimension simply
scales as

\[
\Delta N \propto
  {{{\langle I^2 \rangle} - {\langle I \rangle}^2}\over 
e^2({\langle v^2 \rangle} - {\langle v \rangle}^2)}
\propto {S(V)\over e^2 v^2_{\rm F}}. 
\]
\vskip -2 truemm
}

\vskip -8 truept
\begin{eqnarray}
\Delta N
&~\equiv~&
{\langle ({\rm d} N)^2 \!-\! {\langle {\rm d} N \rangle}^2 \rangle}
%\cr
%\cr
%&=& 
~=~ L{n(\mu_{\rm up})\over 2E_{\rm F}}
{\left[ {\cal T}^2 k_{\rm B}T +
{\cal T}(1 \!-\! {\cal T}){{\mu_{\rm up} \!-\! \mu_{\rm dn}}\over 2}
{\rm coth}
{\left( {{\mu_{\rm up} \!-\! \mu_{\rm dn}}\over 2k_{\rm B}T} \right)}
\right]},
\label{A4}
\end{eqnarray}

\noindent
where $n(\mu_{\rm up})$ is the density in the uniform wire
at equilibrium. Note that LB models work only in
the degenerate limit of large Fermi energy:
$E_{\rm F} \gg k_{\rm B}T$.

Now let the driving field go to zero; the term that depends on
$eV \!=\! \mu_{\rm up} \!-\! \mu_{\rm dn}$ goes to $k_{\rm B}T$.
Using $Ln(\mu) = N$, we are left with the ratio

%\vskip -4 truept
\begin{equation}
{\Delta N\over N}
= {\cal T}{\left[ {\Delta N\over N} \right]}_{\rm eq},
~{\rm where}~
{\left[ {\Delta N\over N} \right]}_{\rm eq}
\equiv {k_{\rm B}T\over 2E_{\rm F}}
= {\kappa\over \kappa_{\rm cl}}.
\label{A5}
\end{equation}

\noindent
Equation (\ref{A5}) very clearly fails to recover the
physical compressibility, Eq. (\ref{A3}), {\it even at equilibrium}.
It depends manifestly on the transport parameter ${\cal T}$.

Momentary reflection shows that Eq. (\ref{A4}), the LB form for
$\Delta N$ -- intimately related to the noise formula -- depends
explicitly on the voltage. Therefore it visits further violence upon the
{\it invariance} of $\kappa$ under transport.

It follows that the Landauer-B\"uttiker noise formula violates
number conservation. The formula is inconsistent with the
compressibility sum rule not only at equilibrium,
which by itself is fatal, but in every transport situation. 

\eject
%\begin{thebibliography}{10}

\end{document}